\documentclass{eas}
\usepackage{graphicx}
\newcommand{\aap}{A\&A}
\newcommand{\araa}{ARAA}
\newcommand{\mnras}{MNRAS}

\newcommand{\apj}{ApJ}

\def \cm2{\mbox{cm$^{-2}$}}
\def \cm3{\mbox{cm$^{-3}$}}

\begin{document}

%%-----------------------------
%%      the top matter
%%-----------------------------
\vspace{-15mm}
\title{Using young massive star clusters to understand star formation and feedback in high-redshift-like environments} 
\runningtitle{Massive clusters: probing high-z star formation}

\author{Steven Longmore}\address{Astrophysics Research Institute, Liverpool John Moores University}
\author{Ashley Barnes$^1$}
\author{Cara Battersby}\address{Harvard-Smithsonian Center for Astrophysics}
\author{John Bally}\address{University of Colorado}
\author{J. M. Diederik Kruijssen}\address{Astronomisches Rechen-Institut, Heidelberg University}
\author{James Dale}\address{Excellence Cluster Universe, Munich}
\author{Jonathan Henshaw$^1$}
\author{Daniel Walker$^1$}
\author{Jill Rathborne}\address{CSIRO Astronomy and Space Science}
\author{Leonardo Testi}\address{European Southern Observatory, Garching bei M\"{u}nchen, Germany}
\author{Juergen Ott}\address{National Radio Astronomy Observatory}
\author{Adam Ginsburg$^7$}

\begin{abstract}

The formation environment of stars in massive stellar clusters is similar to the environment of stars forming in galaxies at a redshift of $1-3$, at the peak star formation rate density of the Universe. As massive clusters are still forming at the present day at a fraction of the distance to high-redshift galaxies they offer an opportunity to understand the processes controlling star formation and feedback in conditions similar to those in which most stars in the Universe formed. Here we describe a system of  massive clusters and their progenitor gas clouds in the centre of the Milky Way, and outline how detailed observations of this system may be able to: (i) help answer some of the fundamental open questions in star formation and (ii) quantify how stellar feedback couples to the surrounding interstellar medium in this high-pressure, high-redshift analogue environment.

\end{abstract}
\maketitle

\section{Massive clusters: a laboratory to study extreme star formation}
\label{sec:laboratory}

Most of what we know about the physics of star formation and stellar feedback is underpinned by observations of star forming regions within the solar neighbourhood, i.e. regions with distances of a few hundred to a few thousand parsecs. The properties of the interstellar medium in the solar neighbourhood are similar to those in the disks of other nearby galaxies, where gas pressures ($P$), surface densities ($\Sigma$) and velocity dispersions ($\sigma$) are typically of the order P/k\,$\sim10^4$\,K\,cm$^{-3}$, $\Sigma\sim100$\,M$_\odot$\,pc$^{-2}$ and $\sigma \sim 10$\,kms$^{-1}$. As such, our understanding of star formation and feedback in the solar neighbourhood is likely a good representation of the star formation across much of the local Universe (Bolatto et al. \cite{bolatto08}).

However, we know that most stars in the Universe formed at a redshift of $\sim1-3$, when the Universe was only a few Gyr old (Madau \& Dickinson \cite{madau_dickinson_2014}). Observations of the interstellar medium in the galaxies forming stars at this epoch show the gas pressures, surface densities and velocity dispersions are factors of several to orders of magnitude larger: P/k\,$\sim10^7$\,K\,cm$^{-3}$, $\Sigma \sim 10^3$\,M$_\odot$\,pc$^{-2}$ and $\sigma \sim 20 - 50$\,kms$^{-1}$(Tacconi et al. \cite{tacconi13}). 

Turbulent star formation theories predict that key end products of the star formation process, such as the stellar initial mass function, as well as the rate and efficiency of star formation, depend sensitively on the local interstellar medium conditions, especially the gas pressure and Mach number. Similarly, cosmological simulations of galaxy formation show that the ultimate fate of a galaxy may be largely determined by the way in which the energy and momentum feedback from young stars couples to the surrounding interstellar medium (Scannapieco et al., \cite{scannapieco12}). This too is highly sensitive to the local interstellar medium conditions. 

Given the predicted strong influence of the local environment on the end products of star formation and feedback, it is clear that our understanding of these processes in the solar neighbourhood is not adequate to describe the conversion of gas into stars at high-redshift. Overcoming this major disconnect in our understanding requires observations to probe the physical processes shaping how individual stars form in these more extreme interstellar medium conditions. However, for the foreseeable future, it will not be possible to resolve individual (forming) stars at the large distances of high-redshift galaxies (Longmore et al., \cite{L14PPVI}). 

Fortunately there are environments in the local Universe with similar interstellar medium conditions to those at high-redshift, and which are also close enough that it is possible to resolve individual (forming) stars within them (Kruijssen \& Longmore, \cite{KL13}). In particular, the precursor clouds to the most massive and dense stellar clusters in the local Universe have gas surface densities, velocity dispersions and pressures of $\Sigma \sim 10^3$\,M$_\odot$\,pc$^{-2}$, $\sigma \sim 20 - 50$ \,kms$^{-1}$, P/k\,$\sim10^7$\,K\,cm$^{-3}$ and can be found within our own Galaxy (Longmore et al., \cite{KL13}).

In this proceeding we describe one such system of gas clouds and young massive clusters, and  outline how detailed observations of this system may be able to (i) help answer some of the fundamental open questions in star formation, and (iii) quantify how stellar feedback couples to the surrounding interstellar medium in this high-pressure environment.

%-------------------------------------------
\section{A system of young massive clusters and progenitor gas clouds}
\label{sec:system}

\begin{figure*}
\begin{center}
\includegraphics[width=0.74\textwidth, angle=0]{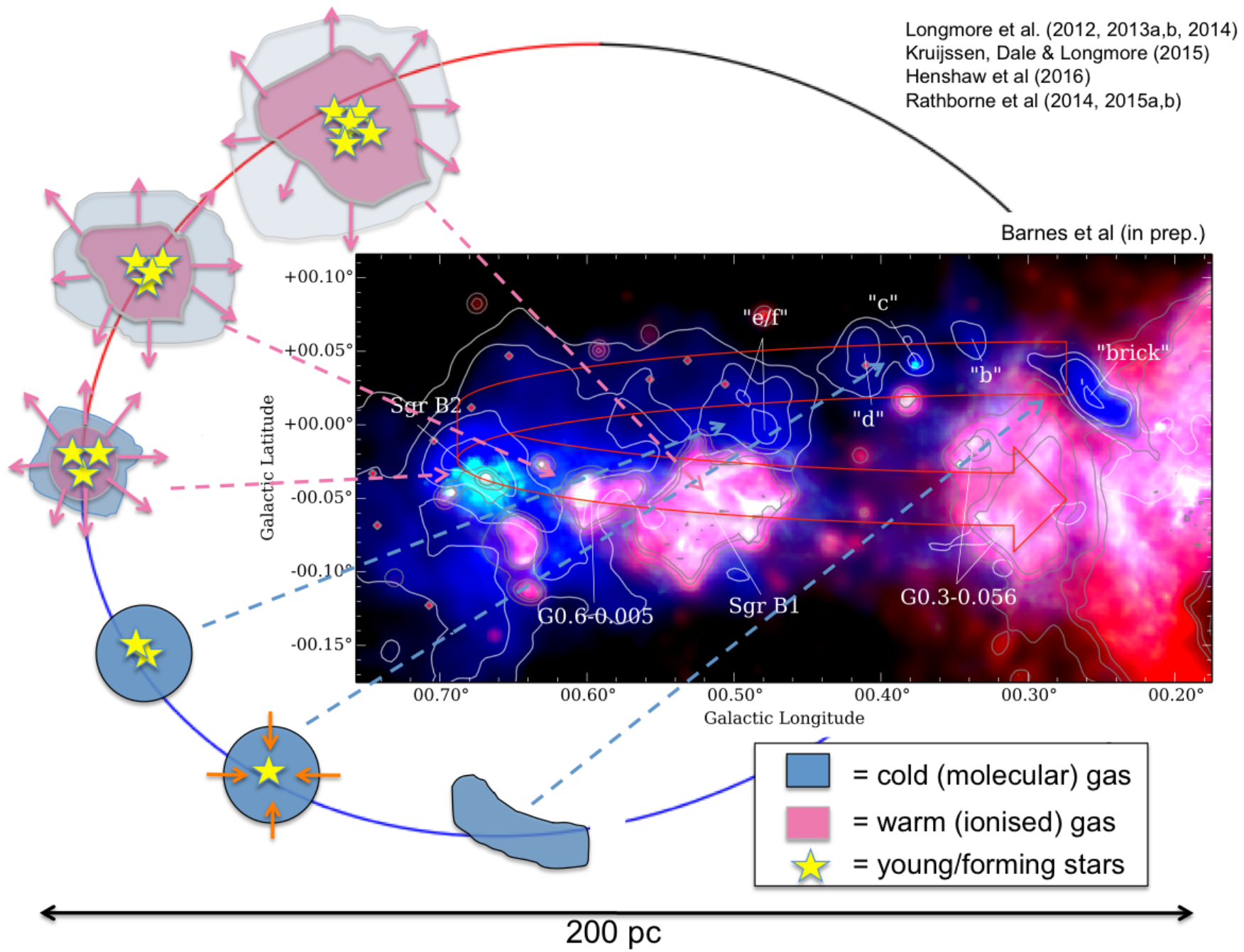}
 \caption{\small  [Inset] Observed protocluster clouds (blue) and young HII regions (red) towards the central $\sim$100\,pc of our Galaxy. [Background/arrows] Schematic diagram of the proposed orbit and relation of clouds in 3D space (Kruijssen, Dale \& Longmore \cite{KDL}).}
 \label{fig:top_down_orbit_plus_data}
 \end{center}
\end{figure*}

The inset on Figure~\ref{fig:top_down_orbit_plus_data} shows observations of the cold molecular gas (blue), and hot ionised gas (red) towards a region within the inner $\sim$100\,pc of our Galaxy (Barnes et al., in prep.).  The white and grey contours show the {\it Herschel} HiGAL column density for the cold and hot gas respectively. This region contains the four most massive and compact (all M$>10^5$\,M$_\odot$, r$\sim$pc-scale) gas clouds in the Galaxy -- the ÒBrickÓ, cloud ÔdÕ, ÔeÕ and ÔfÕ (Immer et al., \cite{immer12}, Longmore et al., \cite{L12}, \cite{longmore13a}). These clouds are clearly linked in position-position-velocity (PPV) space along a coherent gas stream known as the Ôdust ridgeÕ. They show very few signs of active star formation. Following the same coherent PPV structure to larger Galactic longitudes (left on the inset of  Figure~\ref{fig:top_down_orbit_plus_data}) lies the Ômini-starburstÕ protocluster, Sgr B2. At slightly lower latitudes and longitudes lie the HII regions G0.6-0.005, Sgr B1 and G0.3-0.056. Their prodigious cm-continuum flux and bolometric luminosity show these contain a large number of young high mass stars.

\begin{figure*}
\begin{center}
\includegraphics[width=0.74\textwidth, angle=0]{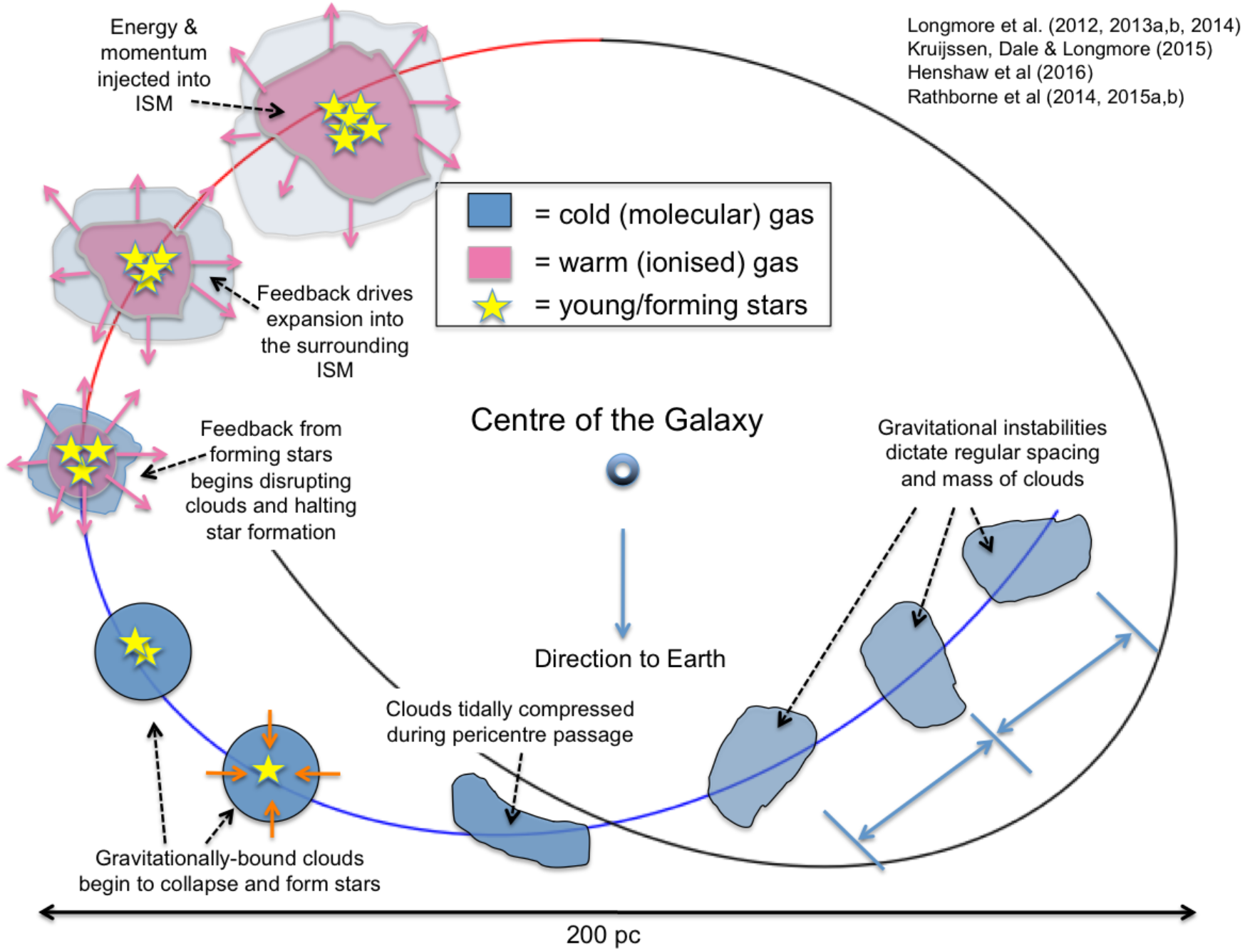}
 \caption{\small  Schematic top-down view of the gas clouds and protoclusters.  }
 \label{fig:top_down_orbit}
  \end{center}
\end{figure*}

Figure~\ref{fig:top_down_orbit} shows a schematic representation of the clouds and HII regions superimposed on a detailed orbital model derived from their kinematics, to show their likely relation in 3D space. The orientation in Figure~\ref{fig:top_down_orbit} is shown as if the clouds and HII regions were viewed looking down from above the plane of the Galaxy, as opposed to the view seen from Earth in the inset of Figure~\ref{fig:top_down_orbit_plus_data}. The arrows in Figure~\ref{fig:top_down_orbit_plus_data} relate the observed edge-on view from Earth with the schematic, top-down view shown in Figure~\ref{fig:top_down_orbit}. 

The interpretation of the 3D structure of the gas clouds and young stellar clusters in Figure~\ref{fig:top_down_orbit} has been derived by comparing the observed kinematics of gas within a galactocentric radius of $\sim$100\,pc, with a parameter space study of possible orbits in the known gravitational potential (Kruijssen, Dale \& Longmore, \cite{KDL}). Any ballistic orbit in the potential is fully defined by 6 free parameters. Thanks to many dense gas surveys of the Galactic centre, the PPV structure of the gas is constrained by many thousands of independent spectra. Using ancillary data (e.g. mid-IR extinction) to determine likely relative distances to different coherent PPV structures, the resulting best-fitting orbit is tightly constrained. In this interpretation, the gas is thought to be directly analogous to circum-nuclear gas streams observed in the centres of most other barred spiral galaxies (Krumholz \& Kruijssen \cite{krumholz_kruijssen}). 

The supermassive black hole, Sgr A*, lies at the bottom of the Galactic gravitational potential and at the focus of the elliptical gas orbit. In this kinematic model the extreme gas clouds and (proto) clusters lie downstream from the gas streams' pericentre passage with Sgr A*, with increasing star formation activity downstream. The gas clouds and protoclusters appear roughly regularly spaced, with projected separations along the stream of $\sim$20\,pc. Intriguingly, the gas clouds upstream from pericentre passage have no star formation activity,  and show evidence of regular, systematic gas flows on roughly the same 20\,pc scale (Henshaw et al., \cite{henshaw16}). Taking the average surface density, $\Sigma \approx 7.5 \times10^2$\,M$_\odot$pc$^{-2}$, and velocity dispersion, $\sigma \approx 6$\,kms$^{-1}$, of gas upstream, the turbulent Jeans length corresponds to $\lambda_J = 2\sigma^2/G\Sigma \sim 22$\,pc. 

Figure~\ref{fig:top_down_orbit} describes the scenario we have put forward to try and explain these observations. In this scenario the gas upstream from Sgr A* is becoming self-gravitating (Kruijssen et al., \cite{kruijssen14}, Krumholz \& Kruijssen \cite{krumholz_kruijssen}), and global gravitational instabilities along the stream lead to the formation of gas condensations separated by the turbulent Jeans length ($\sim$20\,pc: Henshaw et al., 2016). The interaction of the gas in the stream with the strong gravitational potential at pericentre passage compresses the gas perpendicular to the orbit, and strips gas from the cloud which lies beyond the tidal radius at pericentre passage (Rathborne et al., \cite{rathborne14a}, Kruijssen et al., in prep.). In this scenario, the Brick is an example of a cloud currently undergoing strong tidal forces, potentially explaining both the lima-bean-shaped morphology of the cloud and the observed counter-rotation of gas in the cloud with respect to the orbital motion of the stream (Kruijssen et al., in prep.). The tidal compression adds turbulent energy into the gas, but due to the very high gas density ($>10^4\,$cm$^{-3}$), this energy is quickly lost and radiated away through isothermal shocks, triggering star formation (Longmore et al. \cite{longmore13b}). As star formation proceeds, feedback from the young stars will begin disrupting the clouds and halting star formation. Eventually the feedback will be sufficient to drive ionised bubbles into the surrounding interstellar medium, leaving behind an embedded protocluster. 

%-------------------------------------------
\section{Exploiting this system of gas clouds and young massive clusters}
\label{sec:exploiting}

While the possible evolutionary relation between the gas clouds and HII regions is potentially exciting, they also provide many independent snapshots of gas being converted into stars and then dispersing their natal cloud, all within the same global (extreme) environment. This provides a powerful laboratory for understanding star formation/feedback in high-redshift-like conditions, and testing predictions of theoretical models benchmarked against observations of nearby clouds. 

For example, turbulent star formation theories predict that gas pressure is the key variable controlling star formation. Rathborne et al. (\cite{rathborne14b}, \cite{rathborne15}) used ALMA observations of the Brick cloud to show the lack of star formation in the cloud is consistent with the theoretically predicted, environmentally-dependent volume density threshold for star formation. This threshold is orders of magnitude higher than that derived for solar neighbourhood clouds, due to the $\ge$3 orders of magnitude higher pressure. These results provide the first empirical evidence that the current theoretical understanding of molecular cloud structure derived from the solar neighbourhood also holds in high-pressure environments. 

Taking the clouds and HII regions as an ensemble, it is straightforward to determine their relative evolutionary stages from the level of star formation activity. As they all have similar masses/sizes and sit in the same environment, this offers an exciting possibility to isolate evolutionary trends in the gas/stellar properties. For example, by measuring the gas mass structure down to core (0.05\,pc) scales for all regions, one can investigate whether there is a direct mapping in mass from cores to stars by looking for statistically-significant differences in the core mass function with time. Initial work from the SMA Galactic Centre Legacy Survey (Battersby, Keto et al) shows there is a direct trend between the mass of the most massive cores in a region and the level of star formation activity, as would be expected if the cores were gaining mass over time (Walker et al., in prep.). 

Using common approximate age indicators (e.g. masers) and lifetimes of different evolutionary stages, it is possible to place the regions on a rough absolute timeline. With multiple snapshots of clouds/clusters ranging from both ends of the timeline it is then possible to not only determine key global parameters (e.g. the star formation efficiency), but also the spatial and time dependence of those key parameters (e.g. the rate at which gas is converted into stars per free-fall time as a function of radius, or how quickly cores gain mass). With a census of the young stellar population in each region, and noting that the series of ionised bubbles in Figure~\ref{fig:top_down_orbit_plus_data} monotonically increases in radius with star formation activity, it is straightforward to determine the efficiency with which the energy and momentum from the young stars needs to couple to the surrounding ISM in order to drive the expansion of the bubbles (Barnes et al., in prep.). Cosmological simulations show that it is this stellar energy/momentum feedback efficiency in high pressure environments that shapes galaxies, yet these values are currently empirically unconstrained. 

In summary, even considered as completely independent regions, these clouds and young massive clusters have the potential to answer several important open questions in star formation. However, if the clouds and HII regions are part of the same causally-related system described in $\S$\ref{sec:system}, in which star formation is triggered by pericentre passage with Sgr A*, then the absolute star formation timeline is known with unprecedented accuracy. This would open a new frontier in star formation research, moving from the `usual' situation of comparing snapshots of unrelated regions, to one in which we can study the end-to-end physics of star formation and feedback over millions of years as a function of \emph{absolute} time in a coherent time series of molecular clouds.

\end{document}